\documentstyle{article}

\font\tenrm=cmr10
\font\tenit=cmti10
\font\elevenbf=cmbx10 scaled\magstep 1
\font\elevenrm=cmr10 scaled\magstep 1
 1

\font\ninerm=cmr9

\renewenvironment{thebibliography}[1]
 { \elevenrm
   \begin{list}{\arabic{enumi}.}
    {\usecounter{enumi} \setlength{\parsep}{0pt}
     \setlength{\itemsep}{2pt} \settowidth{\labelwidth}{#1.}
     \sloppy
    }}{\end{list}}

\begin{document}
\begin{center}{{\elevenbf
      SEMICLASSICAL QUANTIZATION OF $SU(3)$ SKYRMIONS
\footnote{\ninerm\baselineskip=11pt
The work supported by the Russian Fund for Fundamental Research, grant
95-02-03868a.}
\\}
\vglue 1.0cm
{\tenrm V.B.KOPELIOVICH \\}
{\tenit Institute for Nuclear Research of the Russian Academy of
Sciences,\\ 60th October Anniversary Prospect 7A, Moscow 117312 \\}
\vglue 0.4cm
{\tenrm ABSTRACT}}
\end{center}
\vglue 0.1cm
{\rightskip=3pc
 \leftskip=3pc
 \tenrm\baselineskip=12pt
 \noindent
Semiclassical quantization of the SU(3)-skyrmion zero modes is
performed by means of the collective coordinate method.
The quantization condition known for SU(2) solitons
quantized with SU(3) collective coordinates is generalized for
SU(3) skyrmions with strangeness content different from zero.
Quantization of the dipole-type configuration with large strangeness
content found recently is considered as an example and the spectrum and the
mass splittings of the quantized states are estimated. The energy and baryon
number density of SU(3) skyrmions are presented in the form emphasizing
their symmetry in different SU(2) subgroups of  SU(3), and the lower 
boundary for the static energy of SU(3) skyrmions is derived.
 \vglue 0.2cm}
\vglue 0.1cm
\baselineskip=14pt
\elevenrm
\section{Introduction}
   The chiral soliton approach proposed at first by Skyrme \cite{1} allows 
one to describe the properties of baryons with fairly good 
accuracy \cite{2}-\cite{4}. Considerable progress has been made recently 
also in understanding the properties of few-nucleon systems \cite{5}-\cite{7}. 
Moreover, this approach allows some predictions for the spectrum 
of states
with baryon number $B>1$ \cite{8}-\cite{13}. The quantization of the 
bound states of skyrmions, primarily their zero modes, is a 
necessary step towards realization of this approach. Different aspects
of this problem have been considered, beginning with the papers \cite{2},
\cite{13} and \cite{8,9,14}; however, more general treatment allowing the
consideration of arbitrary $SU(3)$ skyrmions was lacking until recently.

In the sector with $B=2$ besides the $SO(3)$ hedgehog with the lowest
quantum state interpreted as an $H$-dibaryon \cite{8,9}, the $SU(2)$
torus - a bound $B=2$ state - was discovered 10 years ago \cite{15}.
          In the flavor-symmetric $(FS)$ case, when all meson masses in
          the Lagrangian are
          equal to the pion mass, there are three degenerate tori in
          the $(u,d)$, $(d,s)$ and
$(u,s)$ $SU(2)$ subgroups of $SU(3)$. In the flavor symmetry broken
$(FSB)$ case the $(u,s)$ and $(d,s)$ tori are degenerate and heavier
than the $(u,d)$ torus. Another local minimum with large strangeness content
was found recently in the $SU(3)$ extension of the model \cite{16}. This 
configuration is of molecular type and consists of two interacting
$B=1$ skyrmions located in different $SU(2)$ subgroups of $SU(3)$,
$(u,s)$ and $(d,s)$. The attraction between two $B=1$ skyrmions in
optimal orientation which
led to the formation of the torus-like state is not sufficient for this
when both skyrmions are located in different $SU(2)$ subgroups of $SU(3)$
and interact due only to one common degree of freedom. To find this 
configuration a special algorithm was developed allowing for the 
minimization of an energy functional depending on eight functions of three
          variables \cite{16}.
The position of the known $B=2$ classical configurations representing
local minima in $SU(3)$ configuration space is shown on Fig.1 in the 
plane with the scalar strangeness content $C_S$ \cite{17} as $Y$ axis and the 
difference of the $U$- and $D$-contents as $X$-axis. Since the sum of all 
scalar contents is equal to unity, they are defined uniquely at each 
          point of this plot. The $SO(3)$ hedgehog $(1)$ has all contents 
          equal to $1/3$ \cite{12}. Intuitively this is clear, since the 
          basis for the $SO(3)$ solitons is formed by
the matrices $\lambda_2, -\lambda_5, \lambda_7$ and they are located in three
$SU(2)$ subgroups of $SU(3)$ on equal footing. 
The three tori in three different $SU(2)$ subgroups of $SU(3)$ are denoted by
the labels $(2)$, $(3)$ and $(4)$, the $u-d$
symmetric state $(2)$ with $C_S=0$ being of special interest. The 
configurations $(3)$ and $(4)$ can be connected by isorotation in the
$(u,d)$ subgroup. The dipole
type state $(5)$ found recently \cite{16} has a binding energy about half
of that of the torus, i.e., about $0.04$ of the mass of the $B=1$ skyrmion.

The zero modes of solitons have been quantized previously
in a few cases: for $SU(2)$ solitons rotated in the $SU(2)$ 
\cite{2} as well as in the
$SU(3)$ configuration space of collective coordinates
\cite{13,8,14}, and also for $SO(3)$ solitons \cite{8,9}. In the 
case of $SU(2)$ solitons rotated in $SU(3)$ space the quantization 
condition known as the Guadagnini condition \cite{13} was established;
          see also \cite{18}. 

The quantization of the $SU(2)$ $B=1$ hedgehog yields the spectrum of
baryons, mainly the octet and decuplet, and moderate agreement
with the data has been achieved \cite{4}. Quantization of the $SU(2)$ 
          torus in the $SU(3)$ space of collective coordinates
leads to predictions of a rich spectrum of strange dibaryons \cite{19,11}.
Most of them are probably unbound if a natural assumption concerning the
poorly known Casimir energy of the torus-like solitons is made; see also 
the discussion in the last section.

However, these solitons are only particular cases, since other
types of solitons exist, e.g. the above-mentioned solitons of dipole type 
with large strangeness content \cite{16}, point $(5)$ on Fig.1. 
In general, one should expect that the map of the local minima
in the $SU(3)$ configuration space will come more and more complicated with
as the baryon number of the configuration increases. In some
cases the local minima corresponding to larger strangeness content may have
greater binding energy than configurations with small or zero $C_S$.
Therefore, a quantization procedure for arbitrary $SU(3)$ solitons
should be developed. This is a subject of the present paper (\cite{20} 
          contains a preliminary short version).
\section{The Wess-Zumino-Witten term}
 Let us consider the Wess-Zumino (WZ) term in the action which defines
the quantum numbers of the system in the quantization procedure. It was
written by E.Witten in the elegant form \cite{21}:
$$S^{WZ}=\frac{-iN_c}{240\pi^2} \epsilon_{\mu\nu\alpha\beta\gamma}
 \int_{\Omega}Tr\bar{L}_{\mu}\bar{L}_{\nu}\bar{L}_{\alpha}\bar{L}_{\beta}
\bar{L}_{\gamma} d^5x',  \eqno (1)  $$
where $\Omega$ is the 5-dimensional region with  4-dimensional
space-time as its boundary, $N_C$ is the number of colors of the
underlying QCD, and $\bar{L}_{\mu}=U^{\dagger}d_{\mu}U$. As usual, we 
introduce time-dependent collective coordinates for the quantization 
of zero modes according to the relation:
$ U(\vec{r},t)=A(t)U_0(\vec{r})A^{\dagger} (t)$. Integration by
parts is possible then in the expression for the WZ-term in the action,
and we obtain for the WZ-term contribution to the Lagrangian of the system:
$$L^{WZ}=\frac{-iN_c}{48\pi^2} \epsilon_{\alpha\beta\gamma}\int
Tr A^{\dagger}\dot{A} (R_{\alpha}R_{\beta}R_{\gamma} +
L_{\alpha}L_{\beta}L_{\gamma}) d^3x , \eqno (2) $$
where $L_{\alpha}=U_0^{\dagger}d_{\alpha}U_0=iL_{k,\alpha}\lambda_k$ and
$R_{\alpha}=d_{\alpha}U_0 U_0^{\dagger}=U_0 L_{\alpha} U_0^{\dagger}$, or
$$L^{WZ}=\frac{N_c}{24\pi^2}\int \sum_{k=1}^{k=8} \omega_k WZ_k d^3x= 
\sum_{k=1}^{k=8} \omega_kL^{WZ}_k , \eqno (3) $$
with the angular velocities of rotation in the configuration space defined
in the usual way, \\
$A^{\dagger}\dot{A}=-{i \over 2} \omega_k\lambda_k$.
Summation over repeated indices is assumed here and below.
The functions $WZ_k$ can be expressed through the chiral derivatives
$\vec{L}_k$:
$$ WZ_i= WZ_i^R+WZ_i^L=
(R_{ik}(U_0)+\delta_{ik})WZ_k^L, \eqno (4a)  $$
$i,k=1,...8$, and
$$WZ_1^L=-(L_1,L_4L_5+L_6L_7)-(L_2L_3L_8)/\sqrt{3}-
2(L_8,L_4L_7-L_5L_6)/\sqrt{3} $$
$$WZ_2^L=-(L_2,L_4L_5+L_6L_7)-(L_3L_1L_8)/\sqrt{3}-
2(L_8,L_4L_6+L_5L_7)/\sqrt{3} $$
$$WZ_3^L=-(L_3,L_4L_5+L_6L_7)-(L_1L_2L_8)/\sqrt{3}-
2(L_8,L_4L_5-L_6L_7)/\sqrt{3} $$
$$WZ_4^L=-(L_4,L_1L_2-L_6L_7)-(L_3L_5L_8)/\sqrt{3}+
2(\tilde{L}_8,L_1L_7+L_2L_6)/\sqrt{3} $$
$$WZ_5^L=-(L_5,L_1L_2-L_6L_7)+(L_3L_4L_8)/\sqrt{3}-
2(\tilde{L}_8,L_1L_6-L_2L_7)/\sqrt{3} $$
$$WZ_6^L=(L_6,L_1L_2+L_4L_5)+(L_3L_7L_8)/\sqrt{3}-
2(\tilde{\tilde{L}}_8,L_1L_5-L_2L_4)/\sqrt{3} $$
$$WZ_7^L=(L_7,L_1L_2+L_4L_5)-(L_3L_6L_8)/\sqrt{3}+
2(\tilde{\tilde{L}}_8,L_1L_4+L_2L_5)/\sqrt{3} $$
$$WZ_8^L= -\sqrt{3}(L_1L_2L_3)+(L_8L_4L_5)+(L_8L_6L_7), \eqno(5) $$
where $(L_1L_2L_3)$ denotes the mixed product of vectors
$\vec{L}_1$, $\vec{L}_2$, $\vec{L}_3$, etc. and \\
$\tilde{L}_3=(L_3+\sqrt{3}L_8)/2$,
$\tilde{L}_8=(\sqrt{3}L_3-L_8)/2$, $\tilde{\tilde{L}}_3=(-L_3+\sqrt{3}L_8)
/2$, $\tilde{\tilde{L}}_8=(\sqrt{3}L_3+L_8)/2$ are the third and eighth 
          components
of the chiral derivatives in the $(u,s)$ and $(d,s)$ $SU(2)$-subgroups.
Here $[\tilde{\vec{L}}_3\tilde{\vec{L}}_8]=-[\vec{L}_3\vec{L}_8]$, etc.
$R_{ik}(U_0)={1 \over 2}Tr\lambda_iU_0\lambda_kU^{\dagger}_0$ is a real 
          orthogonal matrix, and
$WZ^R_i$ are defined by the expressions $(5)$ with the substitution $\vec{L}_k
          \rightarrow \vec{R}_k$. Relations similar to $(5)$ can be
          obtained for $\widetilde{WZ}_3$ and $\widetilde{WZ}_8$; they are
          analogs
          of $WZ_3$ and $WZ_8$ for the $(u,s)$ or $(d,s)$ $SU(2)$ subgroups,  
      thus clarifying the symmetry of the $WZ$-term in the different $SU(2)$  
        subgroups of $SU(3)$.
          
The baryon number of the $SU(3)$ skyrmions can be written also in terms of
$\vec{L_i}$ in a form where its symmetry in the different $SU(2)$ subgroups
of $SU(3)$ is obvious:
$$ B=-\frac{1}{2\pi^2} \int \biggl((\vec{L_1}\vec{L_2}\vec{L_3}) +
(\vec{L_4}\vec{L_5}\tilde{\vec{L_3}}) +
(\vec{L_6}\vec{L_7}\tilde{\tilde{\vec{L_3}}})+\frac{1}{2}
[(\vec{L_1},\vec{L_4}\vec{L_7}-\vec{L_5}\vec{L_6})+
(\vec{L_2},\vec{L_4}\vec{L_6}+\vec{L_5}\vec{L_7})]\biggr) d^3r. \eqno(6) $$
The contributions of the three $SU(2)$ subgroups enter the baryon number 
          on equal
footing. In addition, mixed terms corresponding to the interaction of
the chiral fields from different subgroups are present also.

It should be noted that the results of calculating the WZ-term according to 
$(5)$ depend on the orientation of the soliton in the $SU(3)$
configuration space.
When solitons are located in the $(u,d)$ $SU(2)$ subgroup of $SU(3)$
only $L_1$, $L_2$ and $L_3$ are different from zero, $WZ_8^R$ and 
$WZ_8^L$ are both proportional to the $B$-number density, and the well known
quantization condition by Guadagnini \cite{13} rederived in \cite{18},
$$ Y_R={2 \over \sqrt{3}}dL^{WZ}/d\omega_8 = N_c B/3,      \eqno(7) $$
applies, where $Y_R$ is the so-called right hypercharge characterizing the
$SU(3)$ irrep under consideration.
This relation is generalized to \cite{20}
$$ Y^{min}_R = {2 \over \sqrt{3}} dL^{WZ}/d\omega_8 = 
N_c B (1 - 3C_S)/3,      \eqno(8)  $$
where the scalar strangeness content $C_S$ is defined in terms of the
real parts of the diagonal matrix elements of the matrix $U$:
$$C_S=\frac{<1-Re U_{33}>}{<3-Re(U_{11}+U_{22}+U_{33})>}, \eqno(9) $$
and $<>$ means averaging or integration over the whole 3-dimensional space
\cite{17}. 
This formula was checked in several cases.

a) One can rotate any $SU(2)$ soliton of the $(u,d)$ subgroup by an arbitrary 
constant $SU(3)$
matrix containing $U_4=exp(-i\nu\lambda_4)$. In this case $C_S=
{1 \over2} sin^2\nu$ \cite{17}, and both $WZ_8^R$, $WZ_8^L$ are proportional
to $R_{88}=1 - {3 \over 2}sin^2\nu$. 
As a result, the relation $(8)$
is fulfilled exactly. Solitons $(3)$ and $(4)$ on Fig.1 can be obtained
from the $(u,d)$ soliton $(2)$ by means of $U_4$ or $U_2U_4$ rotations and
satisfy relation $(8)$. 
For example, when the skyrmion is located in 
the $(u,s)$ $SU(2)$ subgroup of $SU(3)$ we have 
$$L^{WZ}(u,s)=-\frac{\sqrt3 N_CB}{12} (\omega_8-\sqrt{3} \omega_3).
 \eqno(10a) $$
For skyrmions in the $(d,s)$ $SU(2)$ subgroup
$$L^{WZ}(d,s)=-\frac{\sqrt3 N_CB}{12} (\omega_8+\sqrt{3} \omega_3). 
\eqno(10b) $$
Since $C_S=0.5$ in both cases \cite{17}, relation $(8)$ holds.
To derive $(10a,b)$ we have noted that if the soliton is located in any 
          $SU(2)$
subgroup of $SU(3)$ two terms in $(2)$ and $(4a)$ give equal contributions.
 
b) For the $SO(3)$ hedgehog the strangeness content was calculated
previously, $C_S=1/3$, \cite{12,11} and $L^{WZ}=0$ according to \cite{8},
at least for periodic $A(t)$ \cite{9}. The standard assumption that the
angular velocities are constant corresponds to (quasi)periodic behaviour of 
$A(t)$, so,  relation $(8)$ is satisfied.

c) We obtained the relation $(8)$ numerically for solitons of the form
\cite{16}
$$        U = U_L(u,s) U(u,d) U_R(d,s) , \eqno(11) $$ 
with
$        U(u,d)= \exp(ia \lambda_2 ) \exp(ib \lambda_3) $
and $U_L(u,s)$ and $U_R(d,s)$ being deformed interacting $B=1$
$SU(2)$ hedgehogs. For this ansatz we had for the rotated $SU(3)$ Cartan-
Maurer currents \cite{16}:
$$ \begin{array}{ll}
L^r_{1i}=s_ac_al_{3i}, &
L^r_{2i}=d_ia,  \nonumber \\[10pt]
L^r_{3i}=(c_{2a}l_{3i} - r_{3i})/2 + d_ib, & L^r_{4i} 
= c_a l_{1i}, \nonumber \\[10pt]
L^r_{5i}=c_a l_{2i}, & L^r_{6i}=s_al_{1i}+r_{1i}(b), 
\nonumber \\[10pt]
L^r_{7i}=s_a l_{2i}+r_{2i}(b), & L^r_{8i}= 
\sqrt{3} (l_{3i}+r_{3i})/2.
\end{array} \eqno (12) $$
in terms of the $SU(2)$ C-M currents $l_{k,i}$ and $r_{k,i}$ $(i,k=1,2,3)$
and the functions $a$ and $b$, with \\
$r_1(b)=c_br_1-s_br_2$, $r_2(b)=c_br_2+s_br_1$, $c_b=cosb$,
$s_b=sinb$, $c_a=cosa$, etc.\\
$i\vec{l}_k\bar{\tau}_k=U_L^{\dagger}\vec{d}U_L$,
$i\vec{r}_k\tilde{\tau}_k=\vec{d}U_RU_R^{\dagger}$, $k=1,2,3$.
Here $\bar{U}_L(u,s)=f_0+i\bar{\tau}_kf_k$, 
$\tilde{U}_R(d,s)=q_0+i\tilde{\tau}_kq_k$,
$k=1,2,3$, $\bar{\tau}$ and $\tilde{\tau}$ are the Pauli matrices 
corresponding to the $(u,s)$ and $(d,s)$ $SU(2)$ subgroups,
and $f_0^2+...+f_3^2=1$, $q_0^2+...+q_3^2=1$.

$L_i^r=TL_iT^{\dagger}$, 
$U_0 = VT$, $V=U(u,s)exp(ia\lambda_2)$, $T=exp(ib\lambda_3)U(d,s)$.
The chirally invariant quantities, $B$-number density $(6)$, and the second-
order and Skyrme term contributions to the static energy have the same form 
in terms of $L_{ki}$ and $L^r_{ki}$. The formula $(4a)$ should be written
then as
$$ WZ_i=[R_{ik}(V)+R_{ik}(T^{\dagger})] WZ_k^L,  \eqno (4b) $$
with $WZ_k^L$ given in terms of $L^r_n$ according to $(5)$.
In the following we shall omit the index "$^r$" everywhere.
Relations $(10a,b)$ can be checked easily with the help of $(4),(5)$
and $(12)$.

Using $(12)$ and $(5)$ we obtain
$$WZ_8^L=\frac{\sqrt{3}}{2} \bigl((\vec{l}_1\vec{l}_2\vec{l}_3)
+(\vec{r}_1\vec{r}_2\vec{r}_3)+s_a[(\vec{l}_1\vec{r}_2-\vec{r}_1
\vec{l}_2,\vec{l}_3+\vec{r}_3)-(\vec{d}s_a \vec{l}_3,\vec{r}_3-2\vec{d}b)]
\bigr) \eqno(13) $$
It follows from $(13)$ and $(4)$ that at large relative distances, for 
arbitrary but not overlapping solitons, and for $a=0$, we have

$$Y_R^{min}=\frac{2}{\sqrt{3}}L^{WZ}_8=
 \frac{1}{2\sqrt{3}\pi^2} \int WZ_8^L d^3x =\frac{1}{4\pi^2}
\int[(\vec{l}_1\vec{l}_2\vec{l}_3)+(\vec{r}_1\vec{r}_2\vec{r}_3)]d^3x
 = - (B_L+B_R)/2,   \eqno  (14)  $$
where $B_L$ and $B_R$ are the baryon numbers located in the left
$(u,s)$ and right $(d,s)$ $SU(2)$ subgroups of $SU(3)$. 
Relation $(8)$ holds since $C_S=1/2$ for both $(u,s)$ and $(d,s)$
skyrmions.
          Equation $(14)$ does not hold in the general case for
          overlapping solitons, since
there is no conservation law for the components of the Wess-Zumino term.

For the strange skyrmion molecule \cite{16} we should
calculate $(3),(5),(8)$ with $WZ_8=(R_{8k}(V)+R_{k8}(T))WZ_k^L$. The
contribution $-(B_L+B_R)/2$ also appears with some additional
terms which turn out to be small numerically. We obtained
$C_S=0.475$ and $Y_R^{min}=-0.87$ in the $FSB$ case, so relation $(8)$ is 
          satisfied with good accuracy.

It is natural to assume that $(8)$ is valid with good accuracy for any 
$SU(3)$ skyrmions. However, corrections to this relation are not excluded 
by our treatment.
\section{Rotation and static energy}
We start with the well known Lagrangian of the Skyrme model widely used in
the literature since \cite{2}. It depends on the parameters $F_{\pi}=
186 Mev$ (experimental value) and the Skyrme parameter $e$:
$$ L_{Sk}=-\frac{F^2_{\pi}}{16} Tr\bar{L}_{\mu}\bar{L}^{\mu}
+ {1 \over 32e^2} Tr (\bar{L}_{\mu}\bar{L}_{\nu}-\bar{L}_{\nu}\bar{L}_{\mu})^2
+ L_{M} \eqno (15) $$
We take $e=4.12$, close to the value suitable for describing, with a bit 
more complicated Lagrangian, the  mass splittings inside the $SU(3)$ 
multiplets of baryons \cite{4}. The chiral and flavor-symmetry-breaking
mass terms $L_M$ in $(15)$ depending on meson masses will be described in 
detail in Section 4.

 The expression for the rotation energy density of the system depending
on the angular velocities of rotations in the $SU(3)$ collective 
coordinate space defined in Section 2 can be written in more compact 
form than previously \cite{20,16}:
$$L_{rot}=\frac{F_{\pi}^2}{32}\bigl(\tilde{\omega}_1^2+\tilde{\omega}_2^2+
...+\tilde{\omega}_8^2 \bigr)+ $$
$$+{1 \over 16e^2} \Biggl\{
(\vec{s}_{12}+\vec{s}_{45})^2 +
(\vec{s}_{45} +\vec{s}_{67})^2 +
(\vec{s}_{67} -\vec{s}_{12})^2 
 + {1 \over 2} \biggl(
(2\vec{s}_{13}-\vec{s}_{46} -\vec{s}_{57})^2 + 
(2\vec{s}_{23}+\vec{s}_{47} -\vec{s}_{56})^2 + $$
$$+(2\tilde{\vec{s}_{34}}+\vec{s}_{16} -\vec{s}_{27})^2 + 
  (2\tilde{\vec{s}_{35}}+\vec{s}_{17} +\vec{s}_{26})^2 + 
(2\tilde{\vec{s}_{36}} +\vec{s}_{14} +\vec{s}_{25})^2 +
(2\tilde{\vec{s}_{37}} +\vec{s}_{15} -\vec{s}_{24})^2 
\biggr) \Biggr\} \eqno (16) $$

Here $\vec{s}_{ik}=\tilde{\omega}_i \vec{L}_k - \tilde{\omega}_k
\vec{L}_i $, $i,k=1,2...8$ are the $SU(3)$ indices,
and $\tilde{\vec{s}_{34}}=(\vec{s}_{34}+\sqrt{3}\vec{s}_{84})/2$, 
$\tilde{\vec{s}_{35}}=(\vec{s}_{35}+\sqrt{3}\vec{s}_{85})/2$, 
$\tilde{\vec{s}_{36}}=(-\vec{s}_{36}+\sqrt{3}\vec{s}_{86})/2$,
$\tilde{\vec{s}_{37}}=(-\vec{s}_{37}+\sqrt{3}\vec{s}_{87})/2$,
similar to $\tilde{L}_3$ and $\tilde{L}_8$. 

To get $(16)$ we used the identity: $\vec{s}_{ab}\vec{s}_{cd}-
\vec{s}_{ad}\vec{s}_{cb} = \vec{s}_{ac}\vec{s}_{bd}$. The formula $(16)$
possesses remarkable symmetry relative to the different $SU(2)$ 
          subgroups of $SU(3)$.
The functions $L_8$ or $\tilde{L}_8$ do not enter $(16)$ as well as expression
$(6)$ for the baryon number density.
The functions $\tilde{\omega}_i$ are connected with the body-fixed
angular velocities of $SU(3)$ rotations by means of transformation
(see $(8)$ above)
$$ \tilde{\omega}=V^{\dagger}\omega V -T \omega T^{\dagger}, \eqno(17a)$$ 
or
$$ \tilde{\omega}_i=(R_{ik}(V^{\dagger})-R_{ik}(T))\omega_{k}=
R_{ki}\omega_k. \eqno(17b)$$
$R_{ik}(V^{\dagger})=R_{ki}(V)$ and $R_{ik}(T)$ are real orthogonal
matrices, $i,k=1,...8$, and $\tilde{\omega}_i^2 = 2(\omega_i^2 - R_{kl}(U_0)
\omega_k\omega_l)$.
 Expressions for $R_{ik}$ are given in the Appendix for the general case
of the parametrization $(11)$. Relations $(17)$ hold just because we are
operating with rotated functions $L^r_{ki}$ in $(12)$.

The expression for static energy can be obtained from $(16)$ by means of 
the substitution $\tilde{\omega}_i \rightarrow 2L_i$ and 
$\vec{s}_{ik} \rightarrow 2[\vec{L}_i\vec{L}_k]$, \cite{16}.
It can be written in a form which emphasizes quite clearly the lower
boundary for the static energy proportional to the winding (baryon) number
of the system:
$$E_{stat}= \int \Biggl\{\frac{F_{\pi}}{8e}\Biggl[
(\vec{L}_1-2\vec{n}_{23}-\vec{n}_{47}+\vec{n}_{56})^2 +
(\vec{L}_2-2\vec{n}_{31}-\vec{n}_{46}-\vec{n}_{57})^2 +
(\vec{L}_4-2\vec{n}_{53}+\vec{n}_{17}+\vec{n}_{26})^2 +$$
$$ +(\vec{L}_5-2\vec{n}_{34}-\vec{n}_{16}+\vec{n}_{27})^2 +
(\vec{L}_6-2\vec{n}_{73}+\vec{n}_{15}-\vec{n}_{24})^2 +
(\vec{L}_7-2\vec{n}_{36}-\vec{n}_{14}-\vec{n}_{25})^2 +
 {2 \over 9} \biggl(
[\vec{L}_3+\tilde{\vec{L}}_3-{3 \over 2}(\vec{n}_{12}+\vec{n}_{45})]^2+$$ 
$$+[\tilde{\vec{L}}_3+\tilde{\tilde{\vec{L}}}_3 -{3 \over 2}
(\vec{n}_{45}+\vec{n}_{67})]^2 + 
[\tilde{\tilde{\vec{L}}}_3 -\vec{L}_3 -{3 \over 2}
(\vec{n}_{67}-\vec{n}_{12})]^2 \biggr) \Biggr] + M.t. +
3\pi^2 \frac{F_{\pi}}{e}\tilde{B} \Biggr\} d^3\tilde{r}, \eqno (18) $$
where $\tilde{B}$ is the baryon number density given by the integrand 
          in $(6)$, $\tilde{r}=F_{\pi} e r$ and
$\vec{n}_{ik}=[\vec{L}_i\vec{L}_k]$. When $i=4,5$, $k=3$ then 
          $\tilde{\vec{L}}_3$
should be taken in $\vec{n}_{i3}$. For $i=6,7$ $\tilde{\tilde{L}}_3$
should be taken. In $(18)$ we used relations $\tilde{L}_3-\tilde{\tilde{L}}_3
=L_3$ and $(\vec{L}_3+\tilde{\vec{L}}_3)^2+(\tilde{\vec{L}}_3+
\tilde{\tilde{\vec{L}}}_3)^2+(\tilde{\tilde{\vec{L}}}_3-\vec{L}_3)^2=
{9 \over 2}(\vec{L}_3^2+\vec{L}_8^2)$. 
The chiral- and flavor-symmetry-breaking mass term $M.t.$ will be considered
in Section 4.

From $(18)$ we have the inequality
$$ E_{stat}- M.t. \geq 3\pi^2 \frac{F_{\pi}}{e}B.   \eqno(19) $$
This inequality was obtained first by Skyrme \cite{1} for the $SU(2)$ model
and is a particular case of the Bogomol'ny-type bound.

Eight diagonal moments of inertia and $28$ off-diagonal ones define 
          the rotation 
energy, a quadratic form in $\omega_i\omega_k$, according to $(16),(17)$.
The analytical expressions for the moments of inertia are too lengthy
to be reproduced here. Fortunately, it is possible to perform calculations
without explicit analytical formulas, by substituting $(17)$ into $(16)$.

The expression for $E_{rot}$ simplifies considerably when the $(u,d)$ $SU(2)$ 
soliton is quantized in the $SU(3)$ space of collective coordinates:
$$L_{rot}(SU_2) = \frac{F_{\pi}^2}{32}(\tilde{\omega}_1^2+\tilde{\omega}_2^2
+...+\tilde{\omega}_7^2) + \frac{1}{8e^2}\bigl(
\vec{s}_{12}^2+\vec{s}_{23}^2+\vec{s}_{31}^2+{1 \over 4}(\tilde{\omega}_4^2+
...+\tilde{\omega}_7^2)(\vec{l}_1^2+\vec{l}_2^2+\vec{l}_3^2)\bigr), 
\eqno(20a) $$
or
$$L_{rot}(SU_2)=\frac{F_{\pi}^2}{8}\bigl[\vec{\omega}^2\vec{f}^2-
(\vec{\omega}\vec{f})^2+\frac{1-f_0}{2}(\omega_4^2+\omega_5^2+
          \omega_6^2+\omega_7^2)\bigr]
+\frac{1}{8e^2}[\vec{\tilde{\omega}}^2 \vec{l_i}^2 - (\vec{\tilde{\omega}}
\vec{l_i})^2+\frac{1-f_0}{2}\vec{l}_i^2(\omega_4^2+...
          +\omega_7^2)], \eqno(20b) $$
where $\tilde{\omega}_i=[R_{ik}(U_0)-\delta_{ik}]\omega_k=2(f_if_k-\vec{f}^2
\delta_{ik}+f_0\epsilon_{ikl}f_l)\omega_k $ for $i,k=1,2,3$, and 
$\vec{\omega}$ and $\vec{\tilde{\omega}}$ have three components in the $(u,d)$ 
$SU(2)$ subgroup with $\vec{\tilde{\omega}}^2=
4[\vec{\omega}^2\vec{f}^2 - (\vec{\tilde{\omega}} \vec{f})^2]$.
          
          To derive $(20b)$ we used also that
$\tilde{\omega}_4^2+...+\tilde{\omega}_7^2=2(1-f_0)(\omega_4^2+...
 +\omega_7^2)$ .
Here $\vec{l_i}$ parametrizes the chiral derivatives of $U_0$:
$ U^{\dagger}_0 d_k U_0 = i\tau_i l_{i,k} $,
          and the functions $f_0$, $\vec{f}$ define the matrix $U_0$ in
          the usual way. 
$\vec{l}^2_i=(d_if_0)^2+...+(d_if_3)^2$. 
          
 Equation $(20b)$ defines the moments of inertia of arbitrary $SU(2)$ 
          skyrmions rotated in $SU(3)$ configuration space          
 and illustrates well that the $SU(2)$ case is much simpler than general 
 $SU(3)$ case.
The analytical expressions for the moments of inertia of axially symmetric
$SU(2)$ skyrmions, also rotated in $\nu$-direction, can be found 
          in \cite{11,19}.

When the $SU(2)$ hedgehog is quantized in the $SU(3)$ collective 
coordinates space only two different moments of inertia enter \cite{8,13,14}
: $\Theta_1=\Theta_2=\Theta_3$ and $\Theta_4=\Theta_5=\Theta_6=
\Theta_7$. 
For the $SO(3)$ hedgehog the rotation energy also depends
on two different moments of inertia: $\Theta_2=\Theta_5=\Theta_7$ and 
          $\Theta_1=\Theta_3=\Theta_4=\Theta_6=\Theta_8$ \cite{8,9}.
In the case of the strange skyrmion molecule we obtained four different
diagonal moments of inertia \cite{20}: $\Theta_1=\Theta_2=\Theta_N$; 
$\Theta_3$;
$\Theta_4=\Theta_5=\Theta_6=\Theta_7=\Theta_S$ and $\Theta_8$. Numerically
the difference between $\Theta_N$ and $\Theta_3$ is not large while 
          $\Theta_8$ is a bit greater than
$\Theta_S$ (see Table 1). In view of the symmetry properties of the 
configuration many off-diagonal moments of inertia are equal to zero.
Few of them are different from zero, but at least one order of
magnitude smaller than the diagonal moments of inertia, e.g. $\Theta_{46}$
and $\Theta_{57}$. For this reason we shall neglect them here in making
estimates.

The Lagrangian of the system can be written in terms of the angular velocities
of rotation and moments of inertia in the form (in the body-fixed system)
$$ L_{rot}=\frac{\Theta_N}{2}(\omega_1^2+\omega_2^2)+\frac{\Theta_3}{2}
\omega_3^2 +\frac{\Theta_S}{2}(\omega_4^2+\omega_5^2+\omega_6^2+
\omega_7^2)+ \frac{\Theta_8}{2}\omega_8^2 + \Theta_{45}(\omega_4\omega_5-
\omega_6\omega_7) + ...  .\eqno(21) $$
The above-mentioned relations between the different moments of inertia of the 
strange molecule can be obtained in the following way, at large distances 
between the two $B=1$ hedgehogs.
When the $B=1$ skyrmion is located in the $(u,s)$ $SU(2)$ subgroup of $SU(3)$
we obtain from $(12)$ and $(16)$:
$$ L_{rot}(u,s)=\frac{\theta_S}{2}(\omega_1^2+\omega_2^2+\omega_6^2+
\omega_7^2)+\frac{\theta_N}{2}[\omega_4^2+\omega_5^2+\frac{1}{4}(\omega_{3}+
\sqrt{3}\omega_8)^2], \eqno(22a) $$
where we have retained the notations used for the $(u,d)$ $B=1$ soliton.

For the $B=1$ skyrmion in the $(d,s)$ subgroup:
$$L_{rot}(d,s)=\frac{\theta_S}{2}(\omega_1^2+\omega_2^2+\omega_4^2+
\omega_5^2)+\frac{\theta_N}{2}[\omega_6^2+\omega_7^2+\frac{1}{4}(\omega_3-
\sqrt{3}\omega_8)^2], \eqno(22b) $$
with \cite{13,14}
 $$\theta_S={1 \over 8} \int(1-c_F)[F_{\pi}^2+{1 \over
   e^2}(F'^2+2s_F^2/r^2)] d^3r  $$
          $$\theta_N={1 \over 6} \int s_F^2[F_{\pi}^2+{4 \over
          e^2}(F'^2+s_F^2/r^2)] d^3r, \eqno (22c) $$ 
where $F(r)$ is the profile function of the $B=1$ hedgehog and $f_0=cosF$.
Relations $(22c)$ follow immediately from $(20b)$. 
Note, that the combinations of $\omega_3$ and $\omega_8$ which enter the
expressions for the rotation energy $(22a,b)$ and the WZW-term $(10)$ are
orthogonal to each other, as it follows from general arguments.

When two $B=1$ hedgehogs in different subgroups, $(d,s)$ and
$(u,s)$, are located at large distances, we should take the sum of the 
expressions $(22a)$, $(22b)$. Simple relations for the 
$B=2$ moments of inertia $\Theta$ in terms of the $B=1$ inertia $\theta$ 
then appear:

$$\Theta_N = 2\theta_S;  \Theta_S=\theta_N+\theta_S;      
  \Theta_3=\theta_N/2;     \Theta_8=3\theta_N/2=3\Theta_3.   \eqno(23)$$ \\
For interacting hedgehogs in a molecule these relations hold
only approximately (see Table 1 below where some numbers are
corrected in comparison with \cite{20}).

\begin{center}

\begin{tabular}{|l|l|l|l|l|l|l|l|l|l|l|}
\hline
&  $B$  & $M_{cl}$ & $M.t.$ &$C_S$ & $\Theta_N$ & $\Theta_S$ & $\Theta_3$&
$\Theta_8$& $\Theta_{38}$ &$\Theta_{46}=-\Theta_{57}$ \\
\hline
$FS$ & 1 & $1702$ &$46$&--    & $5.55$ &$2.04$ &  -- &  -- &  -- & --  \\
$FS$ & 2 & $3330$ &$87$&$0.495$&$4.14$ &$7.13$ & $2.86$&$8.14$&$0.01$&$0.63$\\
\hline
$FSB$ & 1 &$1982$ & $199$&--    &$3.24$ & $1.06$&  -- &  -- &  -- & -- \\
$FSB$ & 2 &$3885$ & $380$&$0.475$&$2.44$&$4.13$&$1.70$&$4.77$&$0.002$&$0.24$\\
\hline
\end{tabular}
\end{center}
\vspace{1mm}
Table 1. The values of the masses $M_{cl}$, the mass term M.t. (in Mev),
the strangeness content $C_S$ and the moments of inertia (in $10^{-3}$ 
          $Mev^{-1}$) for 
the hedgehog with $B=1$ and the dipole configuration with $B=2$ \cite{16} in 
the flavor-symmetric $(FS)$ and flavor-symmetry-broken $(FSB)$ cases. 
          Here $M.t.$ is 
included in $M_{cl}$,  $F_{\pi}=186$ $Mev$ and $e=4.12$. The accuracy of
          calculations is at least $\sim 0.5 \% $ in the masses and
          few $\%$ in other quantities.

\vglue 0.1in

In the flavor-symmetric $(FS)$ case all meson masses in the Lagrangian
are equal to the pion mass, the distance between centers of both
skyrmions in the molecule equals $\sim 1.05$ $Fm$. In the $FSB$ case the kaon 
mass is included in the Lagrangian (see the
next section) and the distance between solitons centers in the molecule is
$\sim 0.75$ $Fm$ \cite{16,20}.

The Hamiltonian of the system can be obtained by the canonical
quantization procedure \cite{2},\cite{14},\cite{8} which we reproduce
here for completeness. The components of the body-fixed $SU(3)$ angular 
momentum $J^R_k$ can be defined as
$$ J^R_k = dL/ d\omega_k \eqno(24) $$
This definition coincides identically with another one,
$$ J^R_k=\frac{1}{2i} Tr A \lambda_k \pi \eqno(25) $$
where $\pi_{\alpha \beta} = dL/d \dot{A}_{\beta \alpha}$.
In the canonical quantization procedure the substitution 
$$ \pi_{\alpha\beta}=\frac{dL}{d \dot{A}_{\beta \alpha}} \rightarrow 
-i \frac{d}{dA_{\beta\alpha}} \eqno(26) $$
is made. The commutation relations 
$$ [J^R_iJ^R_k] = -if_{ikl}J^R_l \eqno(27) $$ 
then follow immediately, $f_{ikl}$ are the $SU(3)$ structure constants.

After the standard quantization procedure the Hamiltonian of the system,\\
$H=\omega_i dL/d\omega_i - L$, is a bilinear function of the generators
$J^R_i$. For the states belonging to a definite $SU(3)$ irrep the
rotation energy can be written in the simplified form:
$$E_{rot} = \frac{C_2(SU_3) - 3 Y_R^2/4}{2 \Theta_S} +
\frac{N(N+1)}{2} \bigl({1 \over \Theta_N} - {1 \over \Theta_S}
\bigr) + \frac{3 (Y_R - Y^{min}_R)^2}{8 \Theta_8}  \eqno (28)  $$
The second order Casimir operator of the $SU(3)$ group is
$C_2(SU_3)={1 \over 3}(p^2+q^2+pq)+p+q$, $N$ is the right isospin
(see Fig.2) and $p,q$ are the numbers of the upper and low indices
in the tensor describing the $SU(3)$ irrep $(p,q)$. The terms linear in the
angular velocities present in the lagrangian due to the Wess-Zumino-Witten
term are canceled in the Hamiltonian, but they lead to the
quantization condition discussed in the previous section. Corrections of
          the order of $\Theta_{45}^2/\Theta_S^2$ and $(\Theta_N-\Theta_3)
/\Theta_N$ have been neglected in $(28)$. Note that 
$Y_R^{min}$ can take arbitrary noninteger values because it is a
quantity similar to the strangeness content $C_S$ \cite{17}, not a quantum 
number. $Y_R$ is a quantum number and can take only integer values.
The usual spatial angular momentum is $J=0$ here. The correct description of
the usual spatial rotations demands the introduction of a second set 
          of collective
coordinates, as it was done previously \cite{11} for the case of flavor
$SU(2)$. It was shown that the states of lowest energy have $J=0$.

It is clear from expression $(28)$ that for $\Theta_8 \rightarrow 0$ 
the right hypercharge satisfies $Y_R=Y^{min}_R={2 \over \sqrt{3}}L_8^{WZ}$,
otherwise the quantum correction due to $\omega_8$ will be infinite.
For solitons located in $(u,d)$ $SU(2)$ we have $\Theta_8=0$ and 
$Y_R={2 \over \sqrt{3}}L_8^{WZ}=B$ , the 
quantization condition \cite{13,18} with $N_c=3$.

For the skyrmion molecule \cite{16} we have $L^{WZ}_8 \approx
-\sqrt{3}/2$, or $Y^{min}_R \approx -1$, as was explained 
above. The last term in $(25)$ is absent for $Y_R=-1$, and because of
the evident constraints
$$\frac{p+2q}{3} \geq Y_R \geq -\frac{q+2p}{3}  \eqno (29) $$
the following lowest $SU(3)$ multiplets 
are possible: octet, $(p,q)=(1,1)$, decuplet $(3,0)$ and antidecuplet
$(0,3)$, Fig.2. 
The sum of the classical mass of 
the soliton and rotational energy for the $B=2$ octet, $10$ and $\bar{10}$
is equal to $\sim3.53$, $3.74$ and $3.89$ $Gev$ for $Y_R=-1$ (the flavor-
symmetric $FS$-case). The whole $FSB$ mass term described in the
following section, $\Delta M +\delta M_{FS}$, should be added to these 
numbers. When the $FSB$ mass term is
included in the classical mass the sum $M_{cl}+E_{rot}$ equals 
$4.23$, $4.59$ and $4.84$ $Gev$ for the octet, decuplet and $\bar{10}$.
Only the mass splitting part of the mass term, $\delta M_{FSB}$, should be 
added to these numbers (see Table 2 below). The octets with
$Y_R=0$ and $1$ have $M_{cl}+E_{rot}+\Delta M$ equal to $4.61$ and $4.73$ 
          $Gev$ according to $(28)$ (the FS scheme of calculation). The
          $SU(3)$ singlet with
$Y_R=0$ has energy equal to $M_S=M_{cl}+3/(8\Theta_8)$ which,
according to Table 1 equals $\simeq 3.38$ $Gev$ in the $FS$ case. This can be 
compared with the $SO(3)$ hedgehog mass, $M_H=3.272$ $Gev$ for the same 
          values of the parameters \cite{12}.
\section{Mass splitting within $SU(3)$ multiplets of dibaryons}
The mass splittings inside $SU(3)$ multiplets are defined as usual by
the $FSB$ part of the mass terms in the lagrangian density:
$$L_M=\frac{F^2_{\pi}m_{\pi}^2}{16} Tr(U+U^{\dagger}-2) + \frac{F_K^2m_K^2-
F_{\pi}^2 m_{\pi}^2}{24} Tr(1-\sqrt{3}\lambda_8) (U+U^{\dagger}-2)
\eqno (30) $$ 

When $(u,d)$ $SU(2)$ solitons are rotated in the "strange" direction by 
          means of the matrix
$U_4=exp(-i\nu \lambda_4)$,   $(30)$ leads to the substitution
$F^2_{\pi}m_{\pi}^2 \rightarrow F^2_{\pi}m^2_{\pi}+sin^2\nu (F^2_Km^2_K-
F^2_{\pi}m^2_{\pi}) $ \cite{4,11}.
For the ansatz $(11)$, after averaging over all phases in the matrix $A(t)$
except $\nu$, we can rewrite the mass term in the energy density  
in the following form:
$$ M.t.=\frac{F_{\pi}^2m_{\pi}^2}{8}(3-v_1-v_2-v_3)+
  \frac{F_K^2 m_K^2-F_{\pi}^2 m_{\pi}^2}{4}[1-v_3+
(2v_3-v_1-v_2)\frac{sin^2\nu}{2}], \eqno(31a) $$
or, for $F_K=F_{\pi}$
$$ M.t.=\frac{F_{\pi}^2m_{\pi}^2}{4}\bigl[(3-v_1-v_2-v_3) \bigl(1/2+
(m_K^2/m_{\pi}^2 -1 ) C_S \bigr)+ (m_K^2/m_{\pi}^2-1)(2v_3-v_1-v_2)
 \frac{sin^2 \nu}{2}. \bigr]   \eqno(31b) $$

Here $v_1$, $v_2$ and $v_3$ are real parts of the diagonal matrix elements
of the matrix $U$, depending on the functions $f_i$ and $q_i$. 
For the ansatz $(11)$ we have ($b_0=0$) 
$$v_1 = c_{a_0}c_a(c_bf_0-s_bf_3)+s_{a_0}s_ac_b  $$
$$v_2 = c_{a_0}c_a(c_bq_0+s_bq_3) + s_{a_0}s_a[c_b(f_0q_0-f_3q_3)+
s_b(f_3q_0+f_0q_3)] - s_{a_0}(f_1q_1+f_2q_2) $$
$$v_3 = f_0q_0-f_3q_3 + s_a[s_b(f_1q_2-f_2q_1)-c_b(f_1q_1+f_2q_2)]
\eqno(32a) $$ 
$a_0$ and $b_0$ are the asymptotic values of the functions $a,b$. For
the local minimum found recently \cite{16} , $a_0=b_0=0$. In this case 
$(32a)$ simplifies to
$$v_1=c_a(c_bf_0-s_bf_3), $$
$$v_2=c_a(c_bq_0+s_bq_3). \eqno(32b) $$
Here $v_3$ is given by $(32a)$ since it does not depend on $a_0,b_0$.
          If $a_0$, $b_0$ are different from zero the ansatz $(11)$ should
          be written  
 $U=exp(-ia_0\lambda_2)U_L(u,s)U(u,d)U_R(d,s)exp(-ib_0\lambda_3)$ to ensure 
the correct behaviour of $U(\vec{r})$ at $\vec{r} \rightarrow \infty $.
For example, if $a=a_0=\pi/2$, $b=b_0=0$ then $v_1=1$, 
          $v_2=v_3=f_0q_0-f_3q_3-f_1q_1
-f_2q_2$, i.e., the skyrmion is located in the $(d,s)$ $SU(2)$ subgroup.

In the $FS$ case the part of the mass term 
$$ M.t._{FS} = F_{\pi}^2 m_{\pi}^2(3-v_1-v_2-v_3)/8 \eqno (33)$$ 
is included in the classical mass $M_{cl}$ which is
minimized. In the $FSB$ case the second part, 
$$\Delta M = (F_K^2 m_K^2-F_{\pi}^2m_{\pi}^2)(1-v_3)/4 \eqno (34) $$ 
also is included in minimized $M_{cl}$, see Table 1.
In the $FS$ case $\Delta M \simeq 1016 Mev$, while in $FSB$ case 
          it is squeezed $\sim 3$ times.

The mass splitting inside $SU(3)$ multiplets is defined by the term
$$\delta M=-\frac{1}{4}(F_K^2m_K^2 - F_{\pi}^2 m_{\pi}^2)(v_1+v_2-2v_3)
<\frac{1}{2} sin^2\nu>, \eqno(35) $$
which is not included in $M_{cl}$ and is considered as a perturbation
in both cases.
Here $\nu$ is the angle of rotation in the "nonstrange" direction.
For two undeformed hedgehogs at large relative distances we have
$v_1+v_2-2v_3 \rightarrow 2(1-cosF)$ where $F$ is the profile function of 
the $B=1$ hedgehog, and the coefficient of $sin^2\nu$ is the same as for
the rotated $B=1$ $(u,d)$ hedgehog.
Note that in the case of a strange skyrmion molecule with strangeness
content close to $0.5$ the term $(35)$ 
defining the mass splitting within multiplets is negative -
directly opposite to the case when the nonstrange $SU(2)$ solitons are used
as starting configurations and are rotated in the "strange" direction.
The quantity $\delta M$ should be added to the sum of $M_{cl}+E_{rot}$ calculated at the
end of Section 3, and $\Delta M$ should be added in the $FS$ case.

To obtain the mass splitting within $SU(3)$ multiplets we should calculate,
as usual, the matrix elements of the function $<\frac{1}{2}sin^2\nu>=
\frac{1}{3}<1 - D_{88}(\nu)>= \frac{1}{3}(1-I) $ for each component of the
$SU(3)$ multiplets described by the $SU(3)$  $D-$functions. Then the
quantity $I$ is equal to 
$$I=\sum_{\gamma}  C^{8;(p.q);(p.q)_\gamma}_{0,0,0;Y,T,T_3;Y,T,T_3}
C^{8;(p,q);(p,q)_\gamma}_{0,0,0;Y_R,N,M;Y_R,N,M}, \eqno(36) $$
expressed through the Clebsch-Gordan coefficients of the $SU(3)$ group
\cite{22}.
In the case of a strange molecule we have $Y_R=-1$, $N=1/2$ for the octet
and decuplet, $N=3/2$ for $\bar{10}$.
The values of $<\frac{1}{2}sin^2\nu> $ and the mass splittings are 
shown in Table 2.

\begin{center}

\begin{tabular}{|l|l|l|l|l|l|l|l|l|}
\hline
$|p,q;Y,T>$&$I$&$-<sin^2\nu /2>$&$\delta M_{FS}$&$M_{FS}$&$\delta M_{FSB}$&
$M_{FSB}$&$F.st.$&$\epsilon_{FS}$\\
\hline
$|8,1, 1/2> $ & $ -2/10 $ & $-4/10 $ & $-385$ &$4.16$&$-124$&$4.10$&
          $\Lambda N$&$0.14$ \\
$|8,0,1>    $ & $ -1/10 $ & $-11/30$ & $-353$ &$4.19$&$-114$&$4.11$&
          $\Xi N$&$0.15$ \\
$|8,0,0>    $ & $  1/10 $ & $-3/10 $ & $-289$ &$4.26$&$ -93$&$4.13$&
          $\Lambda\Lambda$&$0.14$ \\
$|8,-1, 1/2>$ & $  3/10 $ & $-7/30 $ & $-224$ &$4.32$&$ -73$&$4.15$&
          $\Lambda\Xi$&$0.14$ \\
\hline
$|10,1, 3/2>$ & $-1/8 $   & $-3/8  $ & $-361$ &$4.40$&$-117$&$4.47$&
          $\Sigma N$ & $0.11$ \\
$|10,0, 1>  $ & $ 0   $   & $-1/3  $ & $-320$ &$4.44$&$-104$&$4.48$&
          $\Xi N$   &  $0.11$ \\
$|10,-1, 1/2>$& $ 1/8 $   & $-7/24 $ & $-280$ &$4.48$&$-91 $&$4.50$&
          $\Lambda \Xi$&$0.11$ \\
$|10,-2, 0>  $& $ 1/4 $   & $-1/4  $ & $-240$ &$4.53$&$-78 $&$4.52$&
          $\Xi \Xi$&   $0.10$ \\
\hline
$|\bar{10},2,0>$   & $-1/4$ & $-5/12$& $-401$ &$4.52$&$-130$&$4.72$&
          $N N$&   $0.04$\\
$|\bar{10},1,1/2>$ & $-1/8$ & $-3/8 $& $-361$ &$4.55$&$-117$&$4.72$&
          $\Lambda N$& $0.06$\\
$|\bar{10},0,1>$   & $ 0  $ & $-1/3 $& $-320$ &$4.59$&$-104$&$4.74$&
          $\Xi N$& $0.07$\\
$|\bar{10},-1,3/2>$& $ 1/8$ & $-7/24$ &$-280$ &$4.63$&$-91 $&$4.75$&
          $\Sigma\Xi$& $0.09$\\
\hline
\end{tabular}
\end{center}
\vspace{2mm}
Table 2. The values of $I$, ${1 \over2}sin^2\nu$, the mass splitting
$\delta M$ (in $Mev$) and the masses $M$ (in $Gev$) for the octet, 
decuplet and antidecuplet of dibaryons in the flavor-symmetric and 
flavor-symmetry-broken cases. The binding energy of the configuration
 $\epsilon=(M_1+M_2-M)/(M_1+M_2)$ relative to the final state $F.st.$
is presented. $M_{FS}=M_{cl,FS}+E_{rot,FS}+\Delta M +\delta M_{FS}$,\\    
$M_{FSB}=M_{cl,FSB}+E_{rot,FSB}+\delta M_{FSB}$.          

\vglue 0.1in
For the octet the allowed strangeness of states is $-1,-2,-3$, for the decuplet
it ranges from $-1$ to $-4$, and the nonstrange dibaryons appear in $\bar{10}$,
          $27$-plet, etc. (Fig.2). The masses of the dibaryons calculated
          according to the $FS$ and
          $FSB$ schemes differ, but not very much since the increase of
          the total mass term in the $FS$ case is compensated by the decrease
          of $E_{rot}$ in comparison with the $FSB$ case. The states
          $|10,-2,0>$ and $|\bar{10},2,0>$ are supposed to have $J=1$ and
     the corresponding energy is added, roughly estimated according to our
     previous results \cite{11}.
          
When $FSB$ mass terms are included in the minimized static energy $M_{cl}$ 
they are squeezed by a factor $\sim 3$ due to the smaller 
dimensions of the kaon cloud in comparison with the pion cloud \cite{16},
          therefore, the moments of inertia are greater and $E_{rot}$ is
          smaller in the $FS$
case (see Table 1). The absolute values of the masses  are controlled by the
Casimir energy \cite{23}-\cite{26}, which has the order of magnitude $\sim
- 1 Gev$ for $B=1$ \cite{25,26} and $\sim -2 Gev$ for $B=2$ molecules.  
          
For the $27$-plet the value of the difference of $I$ for states with maximum
and minimum hypercharge is $3/8$, just as for decuplet and antidecuplet.
The relative binding $\epsilon$ is shown in Table 2 because it is less 
sensitive to the method of calculation. $M_1$ and $M_2$ are the masses of 
the final baryons available due to strong interactions, calculated within the 
same approach (theory-to-theory comparison).
Inclusion of configuration mixing usually leads to an increase of
the mass splitting by $\sim 0.3-0.4 $ \cite{27}. Since the results for 
the mass splitting shown in Table 2 depend on the starting configuration,
and both $FS$ and $FSB$ calculation schemes are not consistent by themselves,          
one should use some interpolating procedure, e.g., similar to the slow-
rotator approximation used successfully in \cite{4} for the description 
of the hyperon mass splitting.
\section{Conclusions and discussion}
The quantization scheme for the $SU(3)$ skyrmions has been presented
and the quantization condition known previously \cite{13} is generalized
for skyrmions with arbitrary strangeness content, which allows one to 
investigate the consequences of the existence of different local minima
in $SU(3)$ configuration space. The
quantization condition $(8)$ is valid for all known $B=2$ local minima 
shown in Fig.1. It is proved rigorously
in several cases; in other cases it was confirmed by numerical
calculation. However, some corrections to relation $(8)$ 
cannot be excluded. The moments of inertia 
of arbitrary $SU(3)$ skyrmions can be calculated with the help of 
formulas $(16), (17)$. Both static and rotational energies as well as the 
          baryon 
number density of $SU(3)$ skyrmions are presented in a form which makes 
apparent their symmetry in different $SU(2)$ subgroups of $SU(3)$.

For the dipole-type configuration with $C_S=0.5$ our results are in 
qualitative agreement with those obtained in \cite{28} for the interaction
potential of two strange baryons located at large distances. 
The absolute values of the masses of both $B=1$ and $B=2$ states are 
          controlled by
the Casimir energies, which make a contribution of order $N^0_c$ to
the masses of the configurations \cite{22}-\cite{25}. However, the dipole-type
configuration does not differ much from the $B=2$ configuration within the
product ansatz which we used as a starting point in our calculations
\cite{16}. For this reason the Casimir energy of the dipole can be close to
twice that for the $B=1$ soliton and can cancel in the binding energies of
dibaryons. We conclude therefore that a new
branch of strange dibaryons in addition to those known previously \cite{8,9},
\cite{11} is predicted with a small uncertainty in the absolute values of
masses due to the Casimir energy, relative to the corresponding $B=1$
states. The values of masses and bindings we obtained here cannot,
however, be taken too seriously, not only because the Casimir energy
is poorly known but also because the non-zero mode contributions 
          closely connected with the
Casimir energy (principally the breathing and vibrational modes) have not
          been taken into account. These effects not only decrease the
          binding energies \cite{6,7}, but can make many of the states
          listed in Table 2 unbound.

The prediction by chiral soliton models of a rich spectrum
of baryonic states with different values of strangeness remains one of
the intriguing properties of such models. The comparison with predictions
of the quark or quark-bag models \cite{29,30} is of special interest. 
Some of such models predict the existence of bound strange baryonic states 
\cite{30}, similar to the chiral soliton approach. 

It is difficult to observe these states, especially those which are above 
the threshold for decay due to strong interactions. 
The searches for the H-dibaryon predicted at first within the MIT
quark-bag model \cite{29} have been undertaken in different
experiments, without success till now. It should be noted that 
observation of the H-dibaryon
can be especially difficult by the following reasons.
First, its dimensions are small in the framework of the chiral soliton
approach \cite{12,11}, $R_H\sim 0.5-0.6 Fm$. Therefore, estimates of the 
H-dibaryon production cross section based on the assumption that its
dimensions are close to the dimensions of the deuteron may be too
optimistic. 

Second, it is not clear how 
the transition from an H-dibaryon to two $B=1$ solitons can proceed. 
Schwesinger proposed a nontrivial parametrization allowing for 
the transition from the $SO(3)$ $B=2$ hedgehog to the $B=2$ $SU(2)$ torus
(described in \cite{31}). Within this parametrization the two
configurations are separated by a potential barrier;
moreover, the behaviour of some function in this parametrization is
singular. So, if such a transition is not possible with smooth functions,
it would be difficult to find H-dibaryon in coalescence experiments.
However, further investigations of the predictions of effective field 
theories providing a new approach to the description of the fundamental 
properties of matter are of interest. The near-threshold enhancement
in $p \Lambda$ system which was observed many years ago in, e.g., the reaction
$pp \rightarrow p \Lambda K^+$ \cite{32} and confirmed in recent 
investigations \cite{33} may be a confirmation of soliton
model predictions, because within this approach there is no difference 
between real and virtual levels.

The problem of the H-dibaryon discussed in \cite{8} is that of parity 
doubling: the $SO(3)$ soliton has no definite parity, so a special 
        symmetrization procedure should be done \cite{8}. A similar problem 
exists for the strange molecules also. For the classical configuration of 
molecular type we have different $B=1$ skyrmions in different parts of
space and in different $SU(2)$ subgroups of $SU(3)$. The molecule has 
no definite parity, but these configurations are invariant under the combined
operation of parity transformation and interchange of $SU(2)$ subgroups.
The electric dipole momentum of the molecule is different from zero (this was 
noted by M.Luty). 
(Anti)symmetrization should be performed, similar to the H-particle case,
providing a state of definite parity and removing the e.d.m. of the
quantized state.
  
I am thankful to Bernd Schwesinger for valuable discussions and suggestions
in the initial stages of the work. 
I am indebted also to G.Holzwarth and H.Walliser for their 
interest in the problems of $SU(3)$ skyrmions and useful discussions 
during my visits to Siegen University, and to B.E.Stern for help in the 
          numerical computations.
I appreciate also the support by Volkswagenstiftung, FRG at the 
          beginning of the present work.
\section{Appendix}
Here we sketch the expressions for the matrix elements $R_{ik}$ which
connect the rotation angular velocities in body-fixed and rotated
coordinate systems:
$\tilde{\omega}_i=R_{ki}\omega_k$,\\
 $R_{ik}=R_{ik}(V)-R_{ik}(T^+)$, $R_{ik}(T^+)=R_{ki}(T)={1 \over 2}
Tr\lambda_iT^+\lambda_kT$, $V=U_L(u,s)exp(ia\lambda_2)$, 
$T=exp(ib\lambda_3)U_R(d,s)$, $U_0=VT$. Definitions of $U_L(u,s)$ and 
$U_R(d,s)$ in terms of functions $f_0,...f_3$ and $q_0,...q_3$ are given 
following expression $(12)$.

We use the notations:\\
 $f_{12}^2=f_1^2+f_2^2$,
$q_{12}^2=q_1^2+q_2^2$, $F_1^+=f_0f_1+f_2f_3$, $F_1^-=f_0f_1-f_2f_3$,\\
$F_2^+=f_0f_2+f_1f_3$,
$F_2^-=f_0f_2-f_1f_3$, $F_3^+=f_0f_3+f_1f_2$, $F_3^-=f_0f_3-f_1f_2$,\\
$Q_1^+=q_0q_1+q_2q_3$, $Q_1^-=q_0q_1-q_2q_3$, $Q_2^+=q_0q_2+q_1q_3$,
$Q_2^-=q_0q_2-q_1q_3$,\\
$s_{bb_0}=sin(b-b_0)$, $Q_c=c_bQ_2^-+s_bQ_1^+$, $Q_s=s_bQ_2^--c_bQ_1^+$,\\
$\Delta_F=f_0^2+f_1^2-f_2^2-f_3^2$, $\delta_F=f_0^2-f_1^2+f_2^2-f_3^2$,
$C^+=c_{b+b_0}(q_1^2-q_2^2)-2s_{b+b_0}q_1q_2$,\\
 $S^+=s_{b+b_0}(q_1^2-q_2^2)
+2c_{b+b_0}q_1q_2$, $C^-=c_{bb_0}(q_0^2-q_3^2)+2s_{bb_0}q_0q_3$, $S^-=
s_{bb_0}(q_0^2-q_3^2)-2c_{bb_0}q_0q_3$.    $(A1)$ \\

Here $a_0$ and $b_0$ are asymptotic values of functions $a$ and $b$. For the
strange molecule \cite{16} we have $a_0=b_0=0$. When, e.g., $a=a_0=\pi/2$, 
$b=b_0=0$ hold, the matrix $U$ corresponds to solitons located in the $(d,s)$
$SU(2)$ subgroup of $SU(3)$. \\

$R_{11}=s_{2a_0}s_{2a}(1-f_{12}^2/2)+c_{2a_0}c_{2a}f_0-c_{2b-2b_0}q_0-
s_{2b-2b_0}q_3$ \\
$R_{12}=c_{2a_0}f_3+s_{2b-2b_0}q_0-c_{2b-2b_0}q_3$,
$R_{13}=s_{2a_0}c_{2a}(1-f_{12}^2/2)-c_{2a_0}s_{2a}f_0$,\\
$R_{14}=s_{2a_0}c_aF_2^+-c_{2a_0}s_af_2+c_{2b_0-b}q_2+s_{2b_0-b}q_1$,\\
$R_{15}=-s_{2a_0}c_aF_1^-+c_{2a_0}s_af_1-c_{2b_0-b}q_1+s_{2b_0-b}q_2$,\\
$R_{16}=s_{2a_0}s_aF_2^++c_{2a_0}c_af_2$, $R_{17}=-s_{2a_0}s_aF_1^--c_{2a_0}
c_af_1$, $R_{18}=-\sqrt{3}s_{a_0}c_{a_0}f_{12}^2$. \\
$R_{21}=-c_{2a}f_3-s_{2b-2b_0}q_0+c_{2b-2b_0}q_3$,
$R_{22}=f_0-c_{2b-2b_0}q_0-s_{2b-2b_0}q_3$,\\
$R_{23}=s_{2a}f_3$, $R_{24}=s_af_1+c_{2b_0-b}q_1-s_{2b_0-b}q_2$,\\
$R_{25}=s_a f_2+s_{2b_0-b}q_1+c_{2b_0-b}q_2 $, $R_{26}=-c_af_1$,
$R_{27}=-c_af_2$, $R_{28}=0$.\\

$R_{31}=s_{2a}c_{2a_0}(1-f_{12}^2/2)-c_{2a}s_{2a_0}f_0$,
$R_{32}=-s_{2a_0}f_3$,\\
$R_{33}=c_{2a}c_{2a_0}(1-f_{12}^2/2)+s_{2a}s_{2a_0}f_0-1+q_{12}^2/2$,\\
$R_{34}=c_ac_{2a_0}F_2^++s_as_{2a_0}f_2$,
$R_{35}=-c_ac_{2a_0}F_1^- -s_as_{2a_0}f_1$,\\
$R_{36}=s_ac_{2a_0}F_2^+-c_as_{2a_0}f_2-Q_c$,
$R_{37}=-s_ac_{2a_0}F_1^-+c_as_{2a_0}f_1-Q_s$,\\
$R_{38}=-\frac{\sqrt{3}}{2}(c_{2a_0}f_{12}^2+q_{12}^2)$.\\
$R_{41}=-s_{2a}c_{a_0}F_2^-+s_{a_0}c_{2a}f_2-s_{2b-b_0}q_1-c_{2b-b_0}q_2$,\\
$R_{42}=-s_{a_0}f_1- c_{2b-b_0}q_1+s_{2b-b_0}q_2$,
$R_{43}=-c_{a_0}c_{2a}F_2^--s_{a_0}s_{2a}f_2$,\\
$R_{44}=c_{a_0}c_a\Delta_F+s_{a_0}s_af_0-c_{bb_0}q_0+s_{bb_0}q_3$,\\
$R_{45}=2c_{a_0}c_aF_3^++s_{a_0}s_af_3+s_{bb_0}q_0+c_{bb_0}f_3$,\\
$R_{46}=c_{a_0}s_a\Delta_F - s_{a_0}c_af_0$,
$R_{47}=2c_{a_0}s_aF_3^+-s_{a_0}c_af_3$, $R_{48}=-\sqrt{3}c_{a_0}F_2^-$.\\

$R_{51}=c_{a_0}s_{2a}F_1^+-s_{a_0}c_{2a}f_1+c_{2b-b_0}q_1-s_{2b-b_0}q_2$,\\
$R_{52}=-s_{a_0}f_2-s_{2b-b_0}q_1-c_{2b-b_0}q_2$,
$R_{53}=c_{a_0}c_{2a}F_1^+ +s_{a_0}s_{2a}f_1$,\\
$R_{54}=-2c_{a_0}c_aF_3^--s_{a_0}s_af_3 -s_{bb_0}q_0-c_{bb_0}q_3$,\\
$R_{55}=c_{a_0}c_a\delta_F+s_{a_0}s_af_0 -c_{bb_0}q_0+s_{bb_0}q_3$,\\
$R_{56}=-2c_{a_0}s_aF_3^-+s_{a_0}c_af_3$,
$R_{57}=c_{a_0}s_a\delta_F-s_{a_0}c_af_0$, $R_{58}=\sqrt{3}c_{a_0}F_1^+$.\\
$R_{61}=-s_{a_0}s_{2a}F_2^--c_{a_0}c_{2a}f_2$,
$R_{62}=c_{a_0}f_1$,\\
$R_{63}=-s_{a_0}c_{2a}F_2^-+c_{a_0}s_{2a}f_2+c_{b_0}Q_2^++s_{b_0}Q_1^-$,\\
$R_{64}=s_{a_0}c_a\Delta_F-c_{a_0}s_af_0$,
$R_{65}=2s_{a_0}c_aF_3^+-c_{a_0}s_af_3$,\\
$R_{66}=s_{a_0}s_a\Delta_F +c_{a_0}c_af_0 -C^- -C^+$,
$R_{67}=2s_{a_0}s_aF_3^+ +c_{a_0}c_af_3- S^- -S^+$,\\
$R_{68}=-\sqrt{3}[s_{a_0}F_2^- +c_{b_0}Q_2^++s_{b_0}Q_1^-]$.\\

$R_{71}=s_{a_0}s_{2a}F_1^++c_{a_0}c_{2a}f_1$, $R_{72}=c_{a_0}f_2$,\\
$R_{73}=s_{a_0}c_{2a}F_1^+-c_{a_0}s_{2a}f_1-c_{b_0}Q_1^-+s_bQ_2^+$,\\
$R_{74}=-2s_{a_0}c_aF_3^-+c_{a_0}c_af_3$,
$R_{75}=s_{a_0}c_a\delta_F-c_{a_0}s_af_0$,\\
$R_{76}=-2s_{a_0}s_aF_3^- -c_{a_0}c_af_3+S^- -S^+$,
$R_{77}=s_{a_0}s_a\delta_F+c_{a_0}c_af_0-C^- +C^+$,\\
$R_{78}=\sqrt{3}[s_{a_0}F_1^+-s_{b_0}Q_2^++c_{b_0}Q_1^-]$.\\
$R_{81}=-\frac{\sqrt{3}}{2}s_{2a}f_{12}^2$, $R_{82}=0$,
$R_{83}=-\frac{\sqrt{3}}{2}(c_{2a}f_{12}^2+q_{12}^2)$,
$R_{84}=\sqrt{3}c_aF_2^+$,\\
$R_{85}=-\sqrt{3}c_aF_1^-$, $R_{86}=\sqrt{3}s_a(F_2^++Q_C)$,
$R_{87}=-\sqrt{3}s_a(F_1^--Q_S),
R_{88}=\frac{3}{2}(q_{12}^2- f_{12}^2) $  . $ (A2) $\\

The $R_{8i}$ do not depend on $a_0,b_0$ because the matrices 
          $\lambda_2,\lambda_3$ commute with $\lambda_8$.
The orthogonality of the real matrices $R(V)$ and $R(T)$ can be checked
immediately from these expressions.

\vglue 0.3cm
{\elevenbf\noindent References}
\vglue 0.2cm

\newpage \vglue 1cm
{\elevenbf Figure captions} 
\vglue 1.0cm
Fig.1 Map of the different local minima for classical configurations with
          $B=2$ in the plane $(C_u-C_d)$, $C_S$. Here $C_u$, $C_d$ and
          $C_S$ are the scalar 
quark contents of the soliton, $(1)$ is the $SO(3)$ hedgehog, $(2)$,$(3)$
and $(4)$ are $SU(2)$ tori in the $(u,d)$, $(d,s)$ and $(u,s)$ subgroups
of $SU(3)$, and $(5)$ is the dipole-type configuration (strange skyrmion
molecule). \\
 
Fig.2 $T_3-Y$ -diagrams for the lowest $SU(3)$ multiplets allowed
for the case of $[SU(2)]^3$ configurations, ansatz $(11)$: singlet 
$(p,q)=(0,0)$, octet $(1,1)$, 
decuplet $(3,0)$ and antidecuplet $(0,3)$. The lower dashed line indicates 
isomultiplets with $Y=-1 \simeq Y^{min}_R$, $T=N$; the upper dashed line shows
nonstrange isomultiplets with $Y=B=2$.

\end{document}